\documentclass[%
 reprint,
superscriptaddress,
 amsmath,amssymb,
 aps,
]{revtex4-2}

\usepackage[dvipsnames]{xcolor}
\usepackage{graphicx}
\usepackage{dcolumn}
\usepackage{bm}
\usepackage{MnSymbol}
\usepackage{soul}

\begin{document}

\preprint{APS/123-QED}

\title{Rheology of suspensions of non-Brownian soft spheres across the jamming and viscous-to-inertial transitions}

\author{Franco Tapia}
\altaffiliation[Present Address: ]{Institute of Urban and Industrial Water Management, TU Dresden, Dresden, Germany}
\affiliation{Aix-Marseille Universit\'e, CNRS, IUSTI, Marseille, France}
\author{Chong-Wei Hong}
\affiliation{Aix-Marseille Universit\'e, CNRS, IUSTI, Marseille, France}
\author{Pascale Aussillous}
\affiliation{Aix-Marseille Universit\'e, CNRS, IUSTI, Marseille, France}
\author{\'Elisabeth Guazzelli} 
\affiliation{Universit\'e Paris Cit\'e, CNRS, Mati\`ere et Syst\`emes Complexes (MSC) UMR 7057, Paris, France}

\date{\today}

\begin{abstract}
The rheology of suspensions of non-Brownian soft spheres is studied across jamming but also across the viscous and inertial regimes using a custom pressure- and volume-imposed rheometer. The study shows that the granular rheology found for suspensions of hard spheres can be extended to a soft granular rheology (SGranR) by renormalizing the critical volume fraction and friction coefficient to pressure-dependent values and using the addition of the viscous and inertial stress scales.  This SGranR encompasses rheological behaviors on both sides of the jamming transition, resulting in an approximate collapse of the rheological data into two branches when scaled with the distance to jamming, as observed for soft colloids. This research suggests that suspensions of soft particles across flow regimes can be described by a unified SGranR framework around the jamming transition.
\end{abstract}

\maketitle

Numerous soft-matter systems behave as weak elastic solids at rest and relatively low stresses while acting as liquids above a typical stress known as the yield stress.
These systems are as diverse as colloids, microgels, emulsions, micelles, foams, and granular materials. The remarkable feature of these soft particulate systems is that, while they can jam at high concentration, they can unjam under shear and thus flow beyond the jamming transition (a regime which cannot be accessed for hard grains) owing to the flow-induced elastic deformation of the particles \cite{Sethetal2011,Liuetal2018,Khabazetal2020}. Several experimental \cite{Nordstrometal2010,Paredesetal2013} and numerical \cite{OlssonTeitel2007,Hatano2008,OtsukiHayakawa2009,Tigheetal2010,Kawasakietal2015} studies have used this characteristic to study the critical rheology around the jamming transition in viscous flows of soft particles. By using power-law scalings in the distance to the jamming point, they have demonstrated a critical behavior of the jamming transition with exponents connecting the behavior above and below jamming. More limited (mostly numerical) investigations have been performed in the inertial regime of flow \cite{Chialvoetal2012,Favieretal2017} and a similar critical power-law behavior has been found in soft granular systems \cite{Chialvoetal2012}. 

The objective of this paper is to examine the rheology of soft spherical particles across jamming but also across the viscous and inertial regimes. To this end, we use large hydrogel spheres suspended in fluids of variable viscosity and a custom pressure- and volume-imposed rheometer. We show that this soft particulate system can be described by a soft granular rheology (SGranR) by renormalizing the critical volume fraction and friction coefficient to pressure-dependent values \cite{Kawasakietal2015} and using the superposition of the viscous and inertial stress scales \cite{Trulssonetal2012,Amarsidetal2017,DongTrulsson2020,Tapiaetal2022}. We also demonstrate that the SGranR is capable of capturing the rheological behaviors on both sides of the jamming transition, resulting in an approximate collapse of the rheological data into two branches when scaled with the distance to jamming, as found for soft colloids \cite{Nordstrometal2010}, but now using stress additivity across the flow regimes.

\begin{figure}
\includegraphics[width=0.45\textwidth]{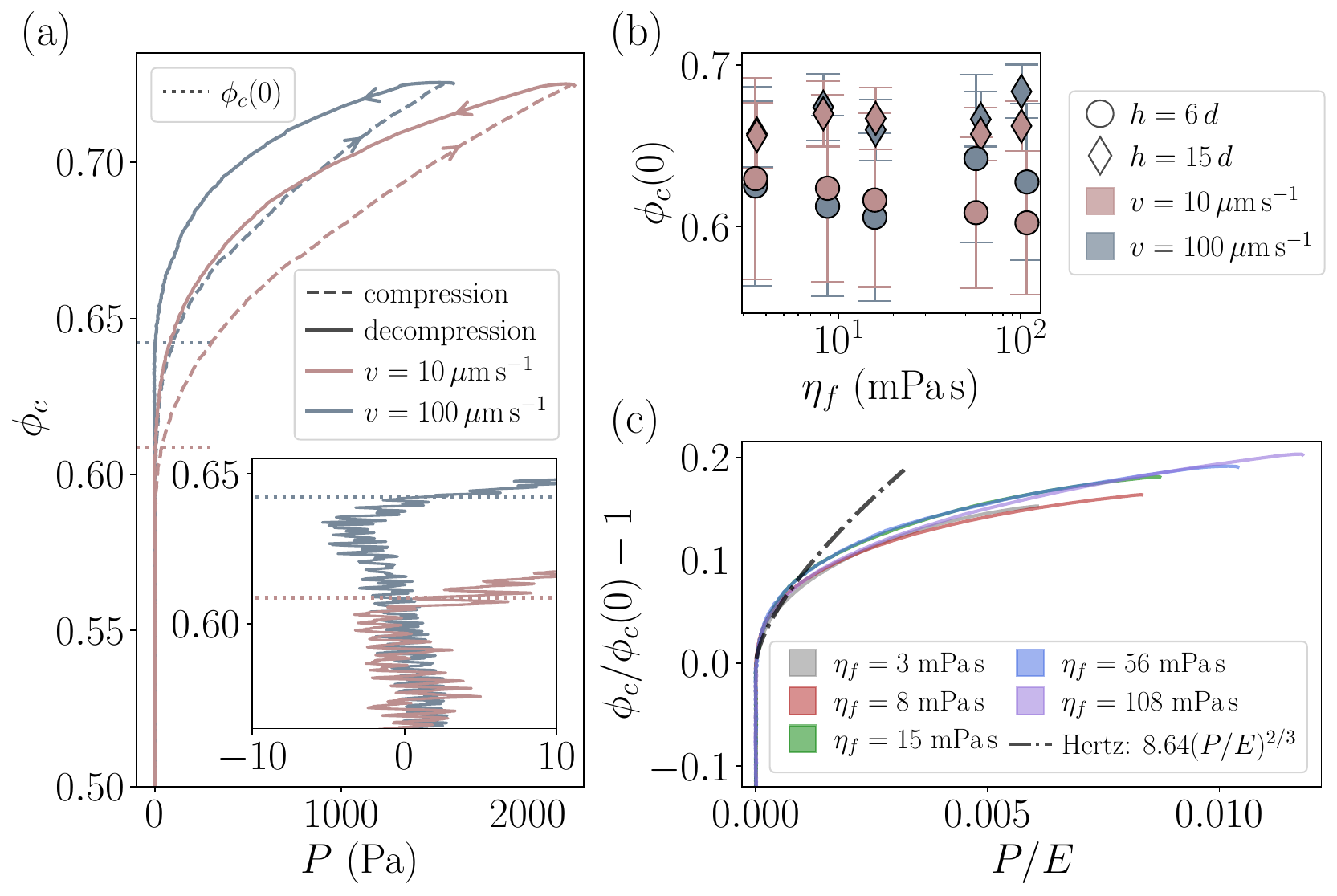}
\caption{\label{fig:Compression} (a) $\phi_c$ versus $P$ for the case $\eta_f = 56$ mPa s and $h = 6 d$ at $v = 10\ \mu$m/s and $= 100 \ \mu$m/s [inset: blowup of the zero pressure zone defining $\phi_c(0)$] during the second cycle of uniaxial compression-decompression, (b) $\phi_c(0)$ (inferred from decompression curves) versus $\eta_f$, and (c) $\phi_c/\phi_c(0)-1$ versus $P/E$ for $h = 6 d$ at $v = 10 \ \mu$m/s from decompression data.}
\end{figure}

The suspensions used in the experiments consist of large polyacrylamide hydrogel spheres grown in a mixture of Ucon oil (Lubricant 75-H-90000) and a solution of $10 \%$ citric acid in water.  Increasing the amount of Ucon oil increases the fluid viscosity $\eta_f$ and leads to a slight decrease in particle diameter $d$ and to an increase in Young's modulus $E$. The hydrogels exhibit low interparticle friction with a coefficient of sliding friction $\mu_{sf} = 0.024\pm 0.004$ when immersed in a 2\% Ucon mixture as measured by a four-ball tester at a rotational speed of 1$s^{-1}$. The measurements of the particle and fluid properties are described in the Supplemental Material 1 \footnote{See Supplemental Material 1 at http:// for the characterization of the particle and fluid properties.}.

\begin{figure*}
\includegraphics[width=0.95\textwidth]{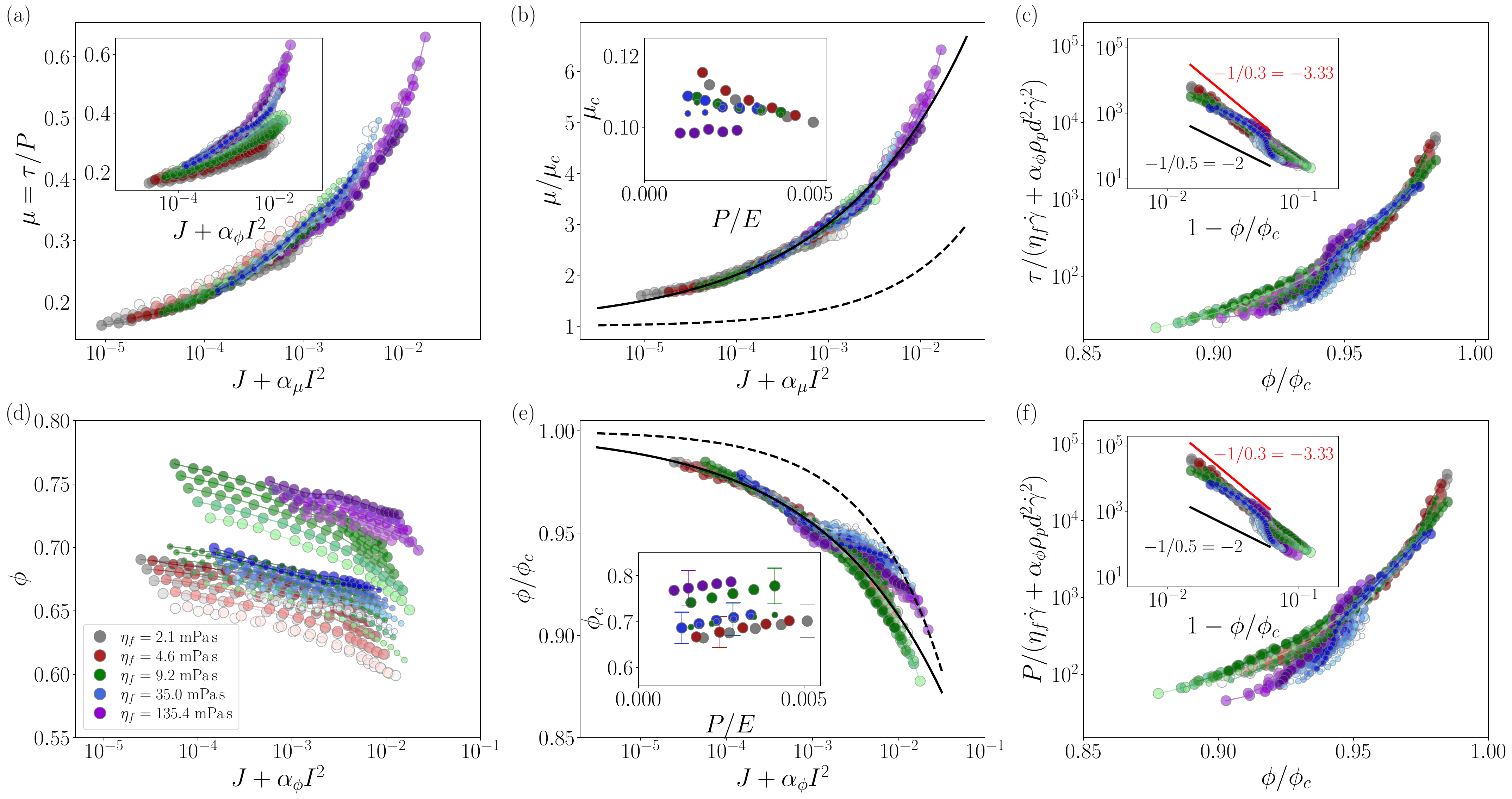}
\caption{\label{fig:P-imposed}Rheological data coming from $P$-imposed rheometry: (a) $\mu=\tau/P$ versus $J+\alpha_{\mu} I^2$ (inset: $\mu$ versus $J+\alpha_{\phi} I^2$) and (d) $\phi$ versus $J+\alpha_{\phi} I^2$; (b) $\mu/\mu_c$ versus $J+\alpha_{\mu} I^2$ (inset: $\mu_c$ versus $P/E$) and (e) $\phi/\phi_c$ versus $J+\alpha_{\phi} I^2$ (inset: $\phi_c$ versus $P/E$); (c) $\tau$ and (f) $P$ normalized by $\eta_f \dot{\gamma} + \alpha_{\phi} \rho_p d^2 \dot{\gamma}^2=\eta_f \dot{\gamma} ( 1+ \alpha_{\phi} St)$  versus $\phi/\phi_c$ (the insets show the same quantities versus the rescaled volume fraction, $1-\phi/\phi_c$).}
\end{figure*}

The mechanical behavior of these hydrogel suspensions are characterized by uniaxial compression-decompression experiments in a (62 mm diameter) cylinder with a porous piston for two different initial heights $h \approx 6$ and $15 d$. Beginning from a loose state obtained by stirring, with the piston just above the particle bed, we impose 2 compression-decompression cycles up to a limit packing fraction $\approx 0.73$ for two piston velocities ($v = 10$ and $100 \ \mu$m/s) and varying fluid viscosity ($3\,\mbox{mPa\,s}\,\lesssim\,\eta_f\,\lesssim\,108\,\mbox{mPa\,s}$) and we focus on the second cycle. 

In Fig.\,\ref{fig:Compression}(a) the packing fraction, $\phi_c$, is plotted as a function of pressure, $P$, for a large $\eta_f = 56$ mPa\,s. We observe a hysteresis as previously reported \cite{Dijksman2013} and a negative pressure at high $v$ (see inset) during decompression due to pore pressure effects, which disappears for low $\eta_f$. We define $\phi_c(0)$ as the position where $\phi_c$ first crosses zero pressure during decompression as indicated in the inset. The reached (non-stationary) values of $\phi_c(0)$ after two cycles are given in Fig.\,\ref{fig:Compression}(b) for the different particle-fluid mixture and are shown to depend mainly on confinement (with smaller values $\approx 0.62 \pm 0.05$ obtained for $h\approx 6 d$ and larger values $\approx 0.66 \pm 0.02$ for $h\approx 15 d$, considering an error of $\pm d/2$ on the height measurement) and slightly on compression rate for large  $\eta_f$ where pore pressure effects occur. Importantly, plotting  $\phi_c/\phi_c(0)-1$ versus the normalized pressure $P/E$ in Fig.\,\ref{fig:Compression}(c) provides a good collapse of all curves, irrespectively of their values of $\phi_c(0)$. A Hertzian behavior \cite{Ohernetal2003} is seen for small strains, followed by a hardening regime previously reported \cite{Lachhab1999, Brodu2015}. The uniaxial compression-decompression data and plots are given in the Supplemental Material 2 \footnote{See Supplemental Material 2 at http:// for the compression data and data analysis.}.

\begingroup
\squeezetable
\begin{table}
    \caption{\label{tab:particlesfluids} Particle and fluid properties of the 5 hydrogel suspensions (at a temperature of $25^{\circ}$ C): percentage of Ucon oil, fluid viscosity $\eta_f$, fluid density $\rho_f$, particle density $\rho_p$, diameter $d$, and Young modulus $E$.}
\begin{ruledtabular}
        \begin{tabular}{ccccccc}
        \%\,Ucon & $\eta_f$\,(mPa\,s) & $\rho_f$\,(kg\,m$^{-3}$) & $\rho_p$\,(kg\,m$^{-3}$) & $d$\,(mm) & $E$\,(kPa)\\ 
        \hline
        2  & 2.1  & 1045 $\pm$ 3 & 1080 $\pm$ 10 & 4.3 $\pm$ 0.1 & 127 $\pm$ 20 \\
        6  & 4.6  & 1050 $\pm$ 3 & 1100 $\pm$ 10 & 4.0 $\pm$ 0.1 & 142 $\pm$ 20 \\
        10 & 9.2  & 1055 $\pm$ 3 & 1110 $\pm$ 10 & 3.8 $\pm$ 0.1 & 157 $\pm$ 20 \\
        20 & 35.0 & 1064 $\pm$ 3 & 1130 $\pm$ 10 & 3.5 $\pm$ 0.1 & 191 $\pm$ 20 \\
        30 & 135.4 & 1074 $\pm$ 3 & 1150 $\pm$ 10 & 3.2 $\pm$ 0.1 & 232 $\pm$ 20 \\
        \end{tabular}
    \end{ruledtabular}
    \end{table}
\endgroup

The rheological responses of 5 hydrogel suspensions of varying fluid viscosity (with physical properties given in Table~\ref{tab:particlesfluids}) are measured with a custom-made annular shear cell \cite{Tapiaetal2019,Tapiaetal2022}. The suspension sample is confined between a lower surface and a porous, planar top plate, within a thickness $24\,\mbox{mm} \lesssim h \lesssim 30\,\mbox{mm}$ (resulting in $h \approx 6-9 d$). The lower surface is rotated at constant velocity producing a linear shear with a shear rate $2\,\mbox{s}^{-1} \lesssim\dot{\gamma}\lesssim 50\,\mbox{s}^{-1}$. The top plate can be moved vertically with a translation stage and enables fluid to flow through it but not particles. Both top and lower surfaces possess a grid trapping the particles which creates a height roughness $\sim 0.5d$. The shear stress $\tau$ is measured from the torque exerted on the top plate after calibration with the pure fluid. The pressure $P$ of the particle phase on the top plate is given by a precision scale attached to the translation stage after correction for buoyancy. The volume fraction $\phi$ is measured from the position of the top plate recorded by a position sensor.  It is important to highlight that this later measurement is contingent upon an accurate estimation of the mass of the particles. This presents a significant challenge for hydrogels, as they must be dried to a specific extent to prevent alterations in their swelling properties. The thin layer of fluid surrounding the spheres may result in an overestimation of $\phi$, particularly for large fluid viscosities. Another source of error comes from accounting of wall roughness for these soft particles. Overall, the error on the absolute value of $\phi$ is estimated to be $\pm 5\%$. The rheometer can be run in $P$-imposed ($250\,\mbox{Pa} \lesssim P \lesssim 650\,\mbox{Pa}$) or $\phi$-imposed ($0.50 \lesssim \phi \lesssim 0.70$) mode using a feedback control loop involving the scale measurement or the position of the top plate, respectively. The rheological data, data analysis, and plots are given in the Supplemental Material 3 \footnote{See Supplemental Material 3 at http:// for the rheological data and data analysis.}. 

We start by presenting the rheological data collected in $P$-imposed rheometry in Fig.\,\ref{fig:P-imposed}. Within this frictional approach, the rheology is given by two dimensionless quantities, the effective friction coefficient $\mu=\tau/P$ and the packing fraction $\phi$. When inertia predominates such as for dry granular media, these quantities depend solely on $I^2= \rho_p d^2 \dot{\gamma}^2/P$ which is the ratio of the inertial stress scale $\sim \rho_p d^2 \dot{\gamma}^2$ and the external confinement pressure $\sim P$ \cite{ForterrePouliquen2008}. Conversely, when viscous forces are dominant, a viscous stress scale $\sim \eta_f  \dot{\gamma}$ is used instead, and the control parameter is $J= \eta_f  \dot{\gamma}/P$ \cite{BoyeretalPRL2011}. An unified theoretical framework across the viscous-to-inertial flow regimes can be established by using superposed viscous and inertial stresses of the form $J+\alpha I^2$ with the coefficient $\alpha$ being the inverse of the Stokes number $St=I^2/J$ at the viscous-inertial transition \cite{Trulssonetal2012,Amarsidetal2017,Tapiaetal2022}.

In Figs.\,\ref{fig:P-imposed}(a)-inset and (d), $\mu$ and $\phi$ are plotted against $J+\alpha_{\phi} I^2$ across the jamming transition for 5 different increasing imposed pressure shown by the increased color intensity for each suspension with varying fluid viscosity. The coefficient $\alpha_{\phi}=0.1$ corresponds to the inverse of the transitional Stokes number $St_{v \rightarrow i} =10$ found for frictional hard spheres \cite{Tapiaetal2022}. In the same way as in this later work \cite{Tapiaetal2022}, a better collapse is obtained for $\mu$ when plotted against $J+\alpha_{\mu} I^2$ with a smaller $\alpha_{\mu}=0.0088$ characteristic of a larger transitional Stokes number for $\mu$, $St^{\mu}_{v \rightarrow i}\approx114$, see the comparison of the main graph and the inset of Fig.\,\ref{fig:P-imposed}(a). In Fig.\,\ref{fig:P-imposed}(e) and (b), the hard-sphere granular rheology \cite{Tapiaetal2022} is extended to the SGranR \cite{Kawasakietal2015},
\begin{eqnarray}
\phi/\phi_c(P/E) & = &  1- a_{\phi} (J+\alpha_{\phi} I^2)^{\gamma_{\phi}},\label{eq:phi}\\
\mu/\mu_c(P/E)  & = & 1+ a_{\mu} (J+\alpha_{\mu} I^2)^{\gamma_{\mu}} \label{eq:mu},
\end{eqnarray}
by normalizing the data by their $P$-dependent critical values at jamming, $\phi_c(P/E)$ and $\mu_c(P/E)$. Fitting the data to Eqs.\,(\ref{eq:phi}-\ref{eq:mu}) (black solid curves) yields $a_{\phi}=0.36\pm0.01$ and $a_{\mu}=16\pm2$, and importantly an exponent $\gamma=\gamma_{\phi}=\gamma_{\mu}=0.3\pm0.1$ differing from the exponent of $0.5$ found for frictional hard spheres \cite{Tapiaetal2022} (black dashed curves) but close to that of $0.35$ predicted for frictionless spheres \cite{DeGiulietal2015}. An impediment to dilation is seen for the most viscous fluids for $J+\alpha_{\phi} I^2 \gtrsim 10^{-3}$ in Fig.\,\ref{fig:P-imposed}(d) and (e) and is likely due to pore pressure effects connected to the negative pressure seen in the bulk decompression experiments depicted in the inset of Fig.\,\ref{fig:Compression}(a). 

\begin{figure}
\includegraphics[width=0.45\textwidth]{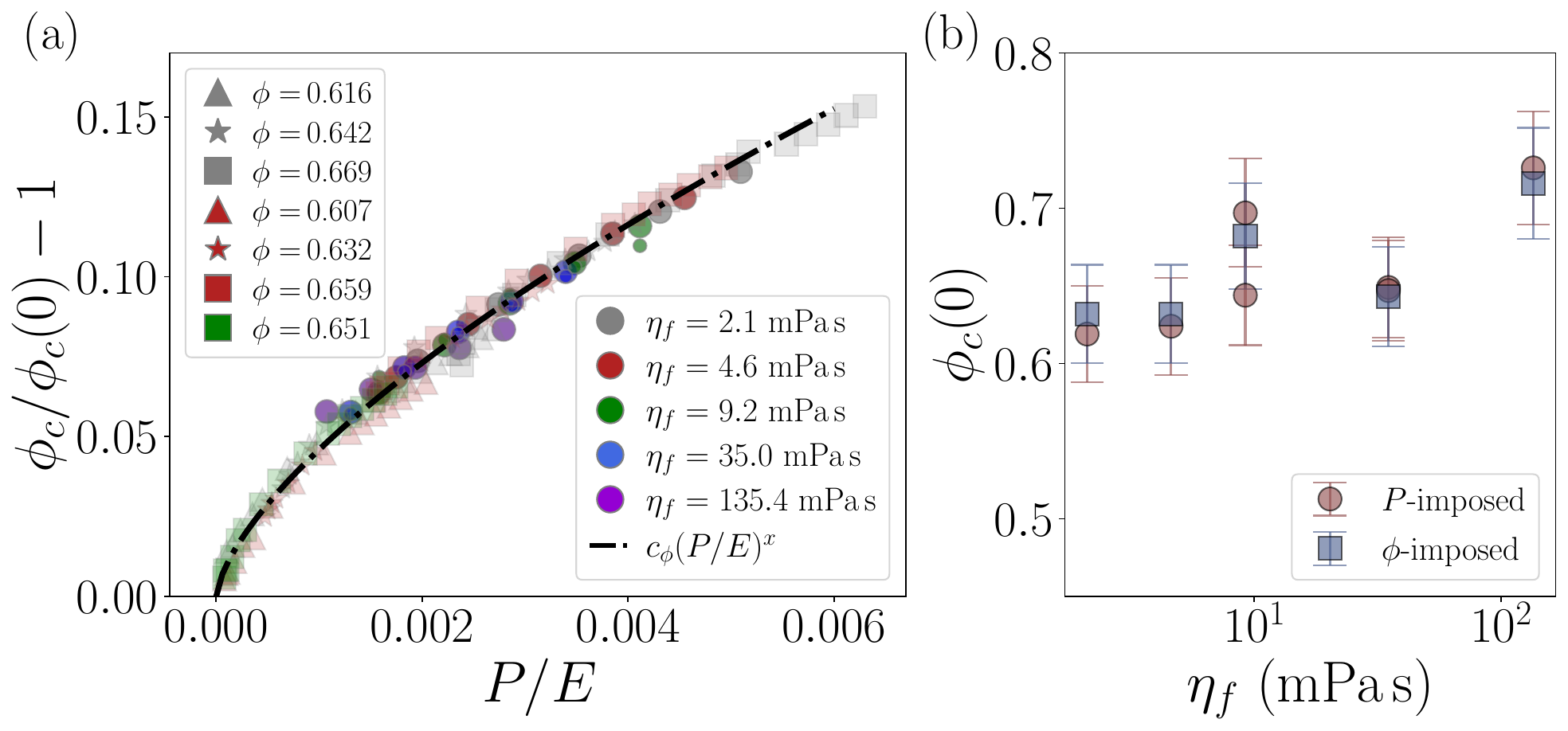}
\caption{\label{fig:phic0}  (a) $\phi_c/\phi_c(0)-1$ versus $P/E$ and (b) $\phi_c(0)$ versus $\eta_f$ for $P$- and $\phi$-imposed data.}
\end{figure}

The quasi-static values, $\phi_c(P/E)$ and $\mu_c(P/E)$, obtained by fitting each imposed-pressure curve using the aforementioned mean values of $a_{\phi}, a_{\mu},$ and $\gamma$, are shown in the insets of Figs.\,\ref{fig:P-imposed}(e) and (b).  Their $P$-dependence can be described by power laws \cite{Kawasakietal2015,Ohernetal2003},
\begin{eqnarray}
\phi_c(P/E)/\phi_c(0)  & = &  1+ c_{\phi} (P/E)^x,\label{eq:phic}\\
\mu_c(P/E)/\mu_c(0) & = & 1- c_{\mu} (P/E)^y \label{eq:muc}.
\end{eqnarray}
While $\mu_c$ shows a weak dependence and has a value $\approx 0.11$ characteristic of low-friction particles, $\phi_c$ presents a conspicuous increase. This latter effect is better evidenced in Fig.\,\ref{fig:phic0}(a) where the relative value $\phi_c/\phi_c(0)-1$ is instead used to avoid the large uncertainties on the absolute value of $\phi_c$. The $P$-imposed data of the inset of Fig.\,\ref{fig:P-imposed}(e) are complemented by data coming from $\phi$-imposed rheometry using Eq.\,(\ref{eq:phi}). Fitting the data to Eq.\,(\ref{eq:phic}) (black dashed curve) yields a coefficient $c_{\phi}=4.7\pm0.2$ and an exponent $x=0.67\pm0.02$ agreeing with Hertz law (for which $x=2/3$) \cite{Ohernetal2003} as obtained at low $P/E$ in the differing experiment of Fig.\,\ref{fig:Compression}(c) which uses uniaxial decompression. The values of $\phi_c(0)$ deduced from the fit are shown in Fig.\,\ref{fig:phic0}(b) for the 5 different suspensions and $P$- and $\phi$-imposed measurements and are close to $\approx 0.64$ within $\pm5\%$, close to the values found in Fig.\,\ref{fig:Compression}(b) under similar confinement.
  
\begin{figure}
\includegraphics[width=0.45\textwidth]{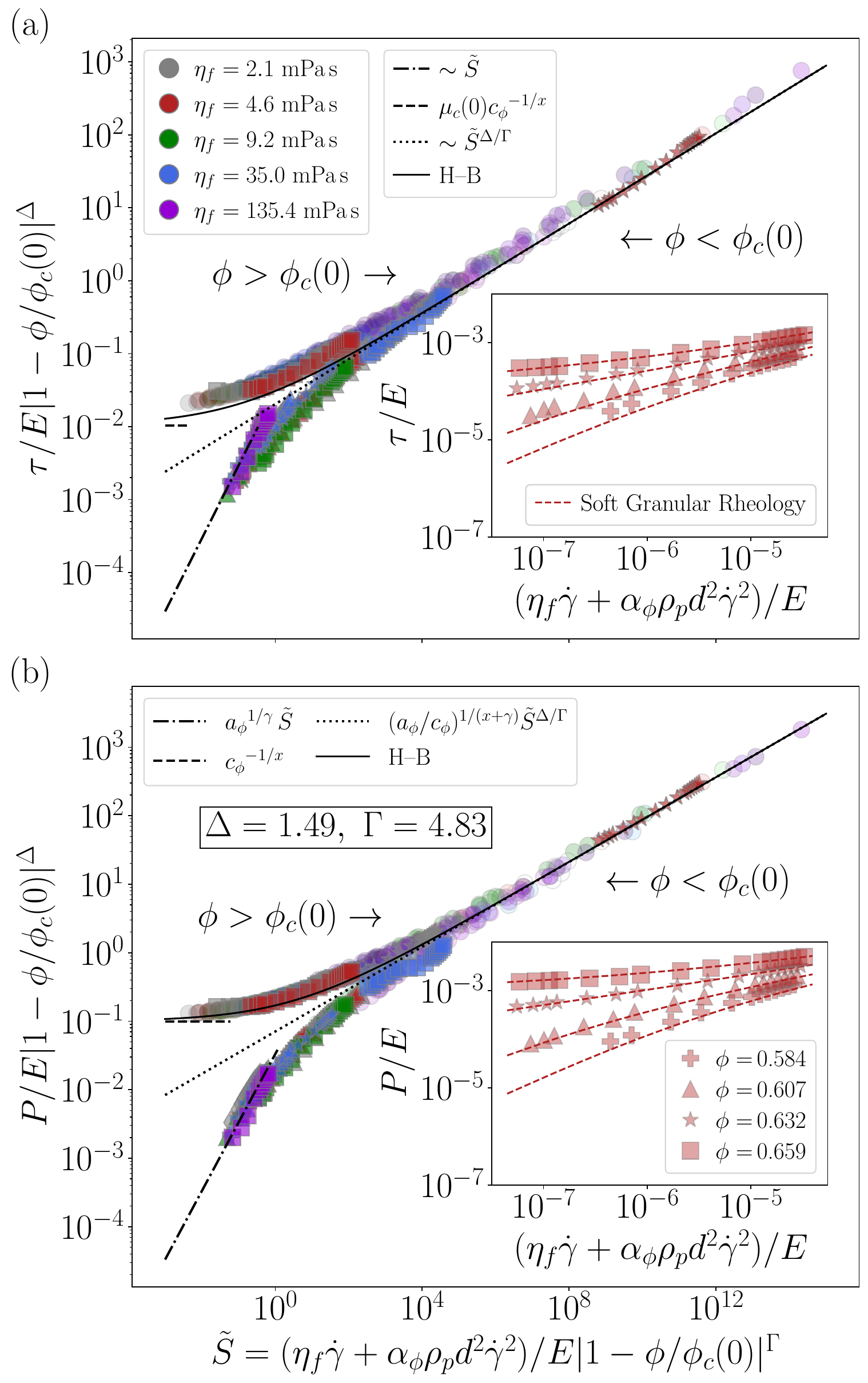}
\caption{\label{fig:collapse} Collapse of the scaled (a) $\tau$ and (b) $P$ against the scaled addition of stress scales with $\tilde S=(\eta_f \dot{\gamma} + \alpha_{\phi} \rho_p d^2 \dot{\gamma}^2)/E |1- \phi/\phi_c(0)|^{\Gamma}$ using the critical exponents $\Delta=1/x$ and $\Gamma=1/x+1/\gamma$, for both $P$- and $\phi$-imposed (for $\phi > \phi_c(0) - 0.1$ to avoid sedimentation effect) data. Insets: (a) $\tau/E$ and (b) $P/E$ versus $(\eta_f \dot{\gamma} + \alpha_{\phi} \rho_p d^2 \dot{\gamma}^2)/E$ for the particle-fluid combination with $\eta_f=4.6$\,mPa\,s.}
\end{figure}

The frictional description of Fig.\,\ref{fig:P-imposed}(b) and (e) has its dual description shown in Fig.\,\ref{fig:P-imposed}(c) and (f), in which the control parameter is $\phi/\phi_c(P/E)$ and the rheology is described by $\tau$ and $P$ normalized by the addition of stress scales $\eta_f \dot{\gamma} + \alpha_{\phi} \rho_p d^2 \dot{\gamma}^2$, using the same $P$-imposed data. The important result shown in the insets is that the normalized $\tau$ and $P$ functions diverge as $(1-\phi/\phi_c)^{-\frac{1}{\gamma}}$ with $\gamma=0.3$ for these low-frictional spheres instead of $(1-\phi/\phi_c)^{-\frac{1}{0.5}}$ for frictional spheres, and thus in better agreement with the theoretical predictions for frictionless spheres \cite{Trulssonetal2012,DeGiulietal2015}.

The SGranR given by Eqs.\,(\ref{eq:phi}-\ref{eq:muc}) provides a description of the rheology of non-Brownian soft particles on both sides of the jamming transition \cite{Kawasakietal2015}, see the full calculation in the Supplemental Material 3 as only the asymptotic limits are reported below. For $\phi > \phi_c(0)$ in the zero shear limit, the SGranR accounts for finite yield pressure and stress (denoted by the subscript $y$),
\begin{eqnarray}
P_y/E  & = & \left\{[\phi/\phi_c(0)-1]/c_{\phi}\right\}^{\frac{1}{x}}, \label{eq:Py}\\
\tau_y/P_y  & = & \mu_c(0) \left\{1 - c_{\mu} \{[\phi/\phi_c(0)-1]/c_{\phi}\}^{\frac{y}{x}} \right\}\approx \mu_c(0). \label{eq:ty}
\end{eqnarray}
For $\phi < \phi_c(0)$ in the zero shear limit, the SGranR rationalizes the diverging behaviors as
\begin{eqnarray}
P/(\eta_f \dot{\gamma} + \alpha_{\phi} \rho_p d^2\dot{\gamma}^2)  & \rightarrow & \left\{[1-\phi/\phi_c(0)] /a_{\phi}\right\}^{-\frac{1}{\gamma}},  \label{eq:PE}\\
\tau/P & \rightarrow & \mu_c(0) \left\{ 1+ \frac{a_{\mu}}{a_{\phi}} \left[1-\phi/\phi_c(0)\right] \right\} \label{eq:tE}.
\end{eqnarray}
Close to jamming, i.e. $\phi \rightarrow \phi_c(0)$, the SGranR produces non-trivial shear-thinning behaviors with, in the present small deformation limit $c_{\phi} (P/E)^x \ll 1$, the approximations
\begin{eqnarray}
P/E  & \rightarrow & \left(a_{\phi}/c_{\phi} \right)^{\frac{1}{\gamma+x}}  \left[(\eta_f \dot{\gamma} + \alpha_{\phi} \rho_p d^2\dot{\gamma}^2)/E\right]^{\frac{\gamma}{\gamma+x}}, \label{eq:PS}\\
\tau/P  & \rightarrow & \mu \left(\phi_c(0), St, \frac{P}{E}\right), \label{eq:tS}
\end{eqnarray}
However, the collapse of the data onto two curves (one above jamming and one below) obtained upon scaling the stresses and shear rate as power laws of the distance to jamming \cite{OlssonTeitel2007} is only approximative within this SGranR framework \cite{Kawasakietal2015}, in particular for the shear stress where the asymptotic behaviors are not connected by simple power laws as seen by Eqs.\,(\ref{eq:ty}),(\ref{eq:tE}), and (\ref{eq:tS}).

Using the critical exponents $\Delta=1/x \,(=1.49$ within $3\%$) and $\Gamma=1/x+1/\gamma \,(=4.83$ within $30\%$) deduced from the asymptotic behaviors given by Eqs.\,(\ref{eq:Py}-\ref{eq:tS}), we plot $\tau/E |1- \phi/\phi_c(0)|^{\Delta}$ and $P/E |1- \phi/\phi_c(0)|^{\Delta}$ versus $\tilde S=(\eta_f \dot{\gamma} + \alpha_{\phi} \rho_p d^2 \dot{\gamma}^2)/E |1- \phi/\phi_c(0)|^{\Gamma}$ using the deduced $\phi_c(0)$ of the different suspensions (given the inset of Fig.\,\ref{fig:phic0}) for both $P$- and $\phi$-imposed data in Fig.\,\ref{fig:collapse}. In the insets, we show typical curves of $\tau$ and $P$ versus $\eta_f \dot{\gamma} + \alpha_{\phi} \rho_p d^2 \dot{\gamma}^2$ (all normalized by $E$) across the jamming transition obtained from $\phi$-imposed measurements which are in excellent agreement with the SGranR predictions (dashed curves). For $\phi > \phi_c(0)$, the stresses extrapolates toward a nonzero yield stress, while for $\phi < \phi_c(0)$, they tend towards zero at low strain rates. In the main plots of Fig.\,\ref{fig:collapse}, a good collapse of the data onto two curves is obtained for the 5 particle-fluid mixtures. As previously mentioned, the collapse happens to be better for the $P$ data than for the $\tau$ data, in particular in the three asymptotic limits. The collapsed branch above $\phi_c(0)$ is approximately a generalized Herschel-Bulkley (H-B) law (solid curve), with yield stress given by Eqs.\,(\ref{eq:Py}-\ref{eq:ty}), and a shear-thinning exponent of $\Delta/ \Gamma\,(=0.3)$, see Eqs.\,(\ref{eq:PS}-\ref{eq:tS}). The collapsed branch below $\phi_c(0)$ is approximately a power law with the same exponent $\approx \Delta/ \Gamma$ which merges with the upper branch very close to the jamming transition (dotted line). Interestingly, this lower branch turns to a linear variation given by Eqs.\,(\ref{eq:PE}-\ref{eq:tE}) (dashed-dotted line) far from jamming as the elastic part of the spheres becomes less dominant.

In conclusion, we have used a custom pressure- and volume-imposed rheometer to obtain reliable rheological data for soft hydrogels above and below the jamming transition and across the viscous and inertial flow regimes. In addition, we have characterized the mechanical properties of these suspensions using uniaxial compression-decompression experiments. The first major result is that we can generalize the granular rheology found for suspensions of hard spheres \cite{Tapiaetal2022}  to a soft granular rheology (SGranR) by renormalizing the critical volume fraction and friction coefficient to pressure-dependent values (weakly-dependent for the friction coefficient) and using the additivity of the viscous and inertial stress scales. The main difference is that the present particles are soft and possess a low friction, which results in their asymptotic behavior differing from that of frictional particles. The second important finding is that this SGranR provides a comprehensive description of the rheological behaviors on both sides of the jamming transition and leads to an approximate collapse of the rheological data into two branches, which is analogous to the behavior observed in soft colloids \cite{Nordstrometal2010,Paredesetal2013}. This is a remarkable result as it means that suspensions of soft particles across the entire range of size and flow, i.e.\,from the colloidal to the granular realm, can be described by the same SGranR framework around the jamming transition.

We greatly thank M. Cloitre for discussions and suggestions. This research is undertaken under the auspices of ANR/DFG Grant No. ANR-21-CE30-0050. It is supported in part by grant NSF PHY-2309135 to the Kavli Institute for Theoretical Physics (KITP).  EG and FT have benefited from the visiting researcher program of the Earthquake Research Institute, University of Tokyo.

\bibliography{BibHydrogels}
\end{document}